\documentstyle[aps,preprint]{revtex}

\tightenlines

\draft

\begin{document}

\title{
Energy Absorption and Storage in a Hamiltonian System
in Partial Contact with a Heat Bath
}
\author{Naoko NAKAGAWA}
\address{
Department of Mathematical Science, Faculty of Science,
Ibaraki University,
Mito, Ibaraki 310-8512, Japan}

\author{Kunihiko KANEKO}
\address{
Department of Pure and Applied Sciences,
College of Arts and Sciences,
University of Tokyo,
Tokyo 153, Japan}

\maketitle

\begin{abstract}

To understand the mechanism allowing for
long-term storage of excess energy in proteins,
we study a Hamiltonian system consisting of several coupled pendula
in partial contact with a heat bath.  
It is found that energy absorption and storage are possible when
the motion of each
pendulum switches between oscillatory (vibrational) and rotational modes. 
The relevance of our mechanism to protein motors is discussed.

\end{abstract}


Proteins are among the most important biopolymers for living systems.
Still, the question of how proteins work is yet unanswered.
Recently a noteworthing experiment concerning protein motors was performed
\cite{Ishijima_Yanagida}.
In this experiment, the working process of a {\it single molecule} 
was directly investigated.
Its results suggest that proteins often store energy obtained 
from a reaction with ATP
and use it for work later (e.g., for enzymatic reactions with other proteins).
The interval for energy storage was found to sometimes be very long, 
up to the order of seconds, while typical time scales for normal vibrations
are several picoseconds. 
How can proteins store excess energy for such a long time,
somehow overcoming the relaxation process
toward thermal equilibrium?
In order for a protein to store energy for a sufficient time, 
energy must be absorbed into a certain part of it,
in accordance with its own dynamics.
Furthermore, some characteristic type of dynamics is required to hold the
excess energy without losing it to the surrounding aqueous solution.

As a first approach to understand the working mechanism of proteins, 
we construct a Hamiltonian system (in partial contact with 
a heat bath), which absorbs and 
stores energy for some time span, in spite of 
its eventual relaxation to thermal equilibrium on a much longer time scale.
In this Letter, we adopt a system consisting of several  
coupled pendula \cite{KK,Antoni_Ruffo}, each of which possesses two modes
of motion, oscillation and rotation. 
We clarify the characteristic dynamics necessary for energy storage
in connection with the coexistence of these two modes, and
also show that partial contact with the heat bath is necessary 
for this storage.
The relevance of our results to protein motors is also discussed.

A protein consists of a folded chain of amino acids that assumes a 
globular shape \cite{Protein_structure}.
Main chains are accompanied by side chains arranged around them,
with each side chain hanging on the main chain as a kind of pendulum.
Some side chains are gathered within a globular shapes,
and a certain assembly of them can play an important role for the
function of the protein.
As an abstract model for the angular motion of
side chains in such a functional assembly, 
we choose a system of coupled pendula.
In particular, we study the idealized case of $N$ identical pendula 
equally coupled to each other.  
Here, the oscillation of the pendula corresponds to the vibration of 
the side chains.
With this simple model as an example, we show that energy absorption 
and long-term storage are generally possible in a class of Hamiltonian systems.
Our study is restricted to a classical mechanical description, since
quantum mechanical effects are believed to be irrelevant to protein dynamics
(except for the choice of the potential).

Our Hamiltonian is given by
\begin{equation}
H=K+V=\sum\limits_{i=1}^N {p_i^2\over 2}+\sum_{i,j=1}^N V(\theta_i,\theta_j),
\end{equation}
where $p_i$ is the momentum of the $i$-th pendulum.
The potential $V$ is constructed so that 
each pair of pendula interacts through their phase difference with an
attractive force to align the phases \cite{KK,Antoni_Ruffo}:
\begin{equation}
V(\theta_i,\theta_j)  =  
{1 \over {2(2\pi)^2N}}\{1-\cos(2\pi(\theta_i-\theta_j))\}.
\end{equation}
Hence the evolution equations for the momentum $p_i$ and the phase 
$\theta_i$ are given by
\begin{eqnarray}
\dot p_j &=& {1\over{2\pi N}}\sum_{i=1}^N \sin(2\pi(\theta_i-\theta_j)),
\label{eqn:seijunP}\\
\dot \theta_j &=& p_j.
\label{eqn:seijunQ}
\end{eqnarray}
In this model, all pairs of pendula interact 
identically.  
If the protein were a straight chain without folding, 
it could be modeled by a one-dimensional chain of pendula.
In reality, due to the globular shape, there is a global interaction
between pendula. Although the assumed ``mean-field coupling" with uniform
strength represents an extreme simplification, the resulting model can 
capture some general characteristics of protein dynamics.

For the (dynamical) function of proteins, it is essential that they exist
in aqueous solution, which plays the role of a heat bath.
Since the hydrophobic part of a protein molecule is
segregated from the aqueous solution, contact with
the heat bath is restricted.
Accordingly, we define the equation of motion so that only some pendula 
in the system contact the heat bath:
\begin{equation}
\dot p_i = -{\partial H \over {\partial \theta_i}} -\gamma p_i
+\sqrt{2\gamma T}\xi_i(t) \qquad (i\leq i_c).
\label{eqn:Langevin}
\end{equation}
Here the heat bath is described by the Langevin equation,
in which $T$ represents its temperature,
$\gamma$ is a relaxation coefficient, and
$\xi_i(t)$ is a Gaussian random form satisfying
$\langle \xi_i(t)\rangle =0$ 
and $\langle \xi_i(t_1)\xi_j(t_2) \rangle
=\delta_{ij}\delta(t_2-t_1)$, with $\langle \cdot \rangle$ 
as the temporal average.
An important point here is that contact with the heat bath is 
restricted only to the few pendula satisfying $i\leq i_c$.
This restriction may be interpreted
as allowing for interaction only of the hydrophilic part
with the aqueous solution.
Note that this restricted heat bath (i.e. partial contact with the heat bath)
is sufficient for realizing thermal equilibrium.

We here briefly review typical behavior of 
Eqs.(\ref{eqn:seijunP}) and ({\ref{eqn:seijunQ}) in the conservative case 
\cite{KK,Antoni_Ruffo,Yamaguchi,Latora_Ruffo}, without coupling 
to the heat bath.
When the total energy is small,
all pendula vibrate almost synchronized
(i.e. $|\theta_i-\theta_j|$ and $|p_i-p_j|$ are small for all $i$, $j$).
The typical time scale of vibration is $\sim 1$ - $10$,
while there is a sort of collective motion, giving rise to
a periodic change of the degree of synchronization, with a
time scale of $\sim 10^2$ - $10^3$.
As the total energy increases, rotational behavior of pendula
begins to appear.
For a sufficiently large total energy,
almost all pendula exhibit rotational motion.
Here, each $p_i$ changes slowly in time, with values of $|p_i-p_j|$ 
relatively large.
In this case the pendula rotate almost freely, 
since the correlation of their phase with those of other pendula
cannot be maintained, and the interaction term cancels out even with 
a short time average,
except in the rare situation that two momenta take very close values.
The time scale of rotation is $O(1)$.

In the medium energy regime,
rotational motion of a single pendulum appears intermittently 
from an assembly of vibrating pendula. Once a pendulum starts to rotate, 
it typically
continues to rotate over many cycles,
which is longer than the typical time scale of each pendulum's 
vibration (and rotation).
We emphasize here that 
the effective interaction for a rotating 
pendulum is {\it much weaker} than those for vibrating ones,
as mentioned above.

When the system size $N$ is very large 
\cite{Antoni_Ruffo,Yamaguchi,Latora_Ruffo},
there appears a phase transition at the energy $E_c/N=0.0190$ ($T_c=0.0127$), 
where the vibrational mode is dominant in the `solid'-like phase and
the free rotation is dominant in the 'gas'-like phase.
Hereafter, we refer to the temperature range around the transition point 
($0.01 \stackrel{<}{\sim} T  \stackrel{<}{\sim} 0.03$)
as the {\it medium temperature} range. 

Now, we consider the Hamiltonian system 
in contact with the heat bath described by Eq.(\ref{eqn:Langevin}).
First, we note that there are two time scales, $\gamma^{-1}$ and $\Gamma^{-1}$,
for the coupling with the heat bath,
where $\Gamma\equiv{\gamma i_c}/N$ gives the dissipation rate of the system.
Below the time scale $\Gamma^{-1}$,
the dynamics are approximated by the Hamiltonian system,
while the dynamics of the pendula corresponding to $i\leq i_c$ are governed
the time scale $\gamma^{-1}$.
On the other hand, for a sufficiently long time scale,
the system arrives at {\it thermal equilibrium},
irrespective of the values of $\gamma$, $i_c \geq 1$, and the temperature.
Note that the canonical ensembles are the same for the $i_c \ll N$ and 
$i_c\simeq N$ cases.

Let us discuss how the dynamics change with $i_c$
by fixing the value of $\Gamma$.
When almost all pendula are in contact with the heat bath,
the time scale to arrive at thermal equilibrium is determined completely by 
the value of $\Gamma$ ($\simeq \gamma$),
independently of the dynamical properties of the system.
On the other hand, in the case of partial contact $i_c\ll N$, 
the time scale to reach thermal equilibrium depends on 
the inherent Hamiltonian dynamics.
For instance, the fluctuations of the total energy 
relax more slowly at both low and high temperature.
In the medium temperature case,
chaos (measured by the maximal Lyapunov exponent) is
stronger \cite{Yamaguchi,Latora_Ruffo}.
This is expected to hasten the relaxation.

In the medium temperature regime, there is a large fluctuation 
of the total energy.
Indeed, the dynamics of the total energy are
correlated with the distribution of energy to each pendulum.  
Here, deviation toward
larger total energy is supported by the concentration of energy
in one (or few) pendulum (see Fig.\ref{fig:eng_abs}) \cite{Takeno_Peyard}.

As the next step, we discuss the energetic behavior of the system 
{\it far from equilibrium}.
Consider a special enzymatic event such as ATP attachment or its reaction.
With such an event, proteins are moved far from equilibrium 
to allow for the accompanying conformational change.  We study how 
such a change is related to the absorption and storage of energy.

Due to the conformational change undergone with such a `reaction' event,
some portions of the side chains are
forced far from the previous equilibrium state.
This situation is modeled by introducing an external potential 
to cause a change in the shape of some part of the system 
as a result of the reaction.
Adopting the coupled pendulum model, we assume that only one pendulum
(to be chosen as the $N$th pendulum) interacts with this external potential.  
Here we choose this external potential as
$V_{N,ext}(\Theta-\theta_N)$, where $\Theta$ is a certain 
phase variable representing the external part.
For the sake of simplicity,
we assume that $(\Theta-\theta_N)$ can be approximately 
written as $\phi_0+\beta \theta_N$ with the
constant $\phi_0$ determining the stable relation between the two parts
and the time scale parameter $\beta$. Choosing a cosine potential as
a simple example, we add the potential term
\begin{equation}
V_{N,ext}=V_0\{1-\cos(2\pi(\beta \theta_N+\phi_0))\},
\end{equation}
where $V_0$ is a constant relating to the strength of the interaction. 

Then, we assume that the `reaction' event occurs at $t=t_0$, thereby changing 
the relation between the pendulum $\theta_N$ and 
the external coordinate $\Theta$.
This event can be accounted for by a
change in the value of $\phi_0$ that forces the pendulum far from equilibrium.
For example, we consider the change from 
$\phi_0=0$ to $\phi_0={1\over 4}$.
Then, after $t_0$,
the equation of motion for the $N$th pendulum is given by
\begin{equation}
\dot p_N= {1\over{2\pi N}}\sum_{i=1}^N \sin(2\pi(\theta_i-\theta_N))
-2\pi\beta V_0 \cos(2\pi\beta \theta_N).
\label{eqn:energy_source}
\end{equation}

An example of time development for $t>t_0$ is shown in Fig.\ref{fig:eng_str}.
Here, $t_0$ is chosen so that $\theta_N$ equals $0$ when 
the $N$th pendulum is suddenly put in a high energy state parameterized 
by $V_0$.
In the computation we used a system with $N=10$ and $i_c=1$;
i.e. only the first pendulum is in contact with the heat bath.

Through the interaction among the pendula,
one (or few) pendulum absorbs 
a large portion of energy in Fig.\ref{fig:eng_str}.
This pendulum, after the relaxation of the $N$th pendulum,
continues to rotate in isolation with a large amount of energy.
Then, it is affected very little by the other pendula 
over a long time interval.
This fact allows for long-term energy storage.  
The reason that such a phenomenon exists is that the interaction 
experienced by the pendulum with such high energy
is quite weak, as mentioned above.

We define the lifetime for energy storage as
the interval between the relaxation time of the $N$th pendulum
and that of the total energy (see Fig.\ref{fig:eng_str}).
The latter is determined
by the time $t_R$ at which the total kinetic energy $K$
decreases to $NT/2$, i.e., the value at thermal equilibrium.
On the other hand, the relaxation time of the $N$th pendulum is tentatively
defined as the time $t_{R0}$ at which $p_N^2/2+V_{N,ext}/N$ decreases to $T$.
The results, however, do not depend on the specific choices of
these relaxation times.

The distribution of the lifetime is shown in Fig.\ref{fig:lifetime}(a),
where $200$ random sequences of $\xi_i(t)$ are sampled.
For comparison, we show the distribution for two cases
with the same value of $\Gamma$: $i_c=1$ with $\gamma=10^{-2}$ 
and $i_c=N(=10)$ with $\gamma=10^{-3}$.
It is noted that the distribution of the lifetime is quite different 
for the two cases.
For the case of partial contact, $i_c=1$,
the typical lifetime of the energy storage is very long 
and reaches $10^7$.
This contrasts with values of approximately $10^2$ - $10^3$ for the case
with full contact, $i_c=N$.
This result suggests that {\it partial contact with heat bath 
is necessary for long-term energy storage.}

Although the thermal equilibrium properties are common in the above two cases,
a large difference appears when the system is placed far from equilibrium.
The pendula corresponding to $i\le i_c$, interacting directly with 
the heat bath,
cannot rotate freely over the $1/\gamma$ time scale
(in contrast with the Hamiltonian dynamics)
and the dynamics are replaced by Brownian motion due to the heat bath.
In the case of partial contact, 
a long duration of rotation is possible for pendula corresponding to $i>i_c$
only if the pendula possess sufficiently large energy
to remain far from equilibrium.
The pendula there become free from the thermal effect
communicated through the distinctively weaker interaction with 
the other pendula,
and nearly follow purely Hamiltonian dynamics.
Although the prototype of this
mechanism is observed as a large fluctuation around equilibrium
in the medium temperature regime,
it works clearly far from equilibrium.

The temperature dependence of the mean lifetime of energy storage 
is displayed in Fig.\ref{fig:lifetime}(b).
For $i_c=N$, the mean lifetime is found to be almost the same 
for every temperature. 
Here, the distribution of lifetimes itself depends only on 
the temperature. The relaxation process is determined
by the heat bath as in the case of thermal equilibrium, 
independently of the temperature.
On the other hand, for $i_c=1$, 
the relaxation time depends strongly on the detailed dynamical 
properties of the system at each temperature.

The mean lifetime of energy storage has a maximum 
around $T=0.01$ - $0.02$.  In this temperature range,
the dynamics are most strongly chaotic, leading to 
rapid absorption of energy, 
and to the coexistence of rotational and vibrational modes,
which is  essential for energy storage.
Also, the mixing properties of chaos are expected to determine the ability 
to transfer the excess energy localized at the $N$th pendulum.
A larger absorption of energy leads to a longer relaxation time.

Energy absorption and storage in the case of
partial contact is observed for a sufficiently large $V_0$, 
small $\gamma$, and within a moderate range of $\beta$ 
($10^{-3} \stackrel{<}{\sim} \beta  \stackrel{<}{\sim} 10^{-2} $).
$V_0$ must be large in order to supply the system enough  
energy to kick it  far from equilibrium.
Unless $\gamma$ is small, the characteristic features of the 
Hamiltonian dynamics are largely destroyed by contact with the heat bath, 
and the ability of the system to absorb energy from the $N$th pendulum is
greatly weakened.
If $\beta$ is out of the above range,
the time scale characterizing the motion of the $N$th pendulum 
is much different from that of other pendula,
and for this reason the other pendula effectively cannot interact 
with the $N$th pendulum and they cannot absorb its energy.

What type of Hamiltonian is required for long-term energy storage?
Consider a system near the region of a phase transition.
Here, some elements possessing higher energy ``melt" 
and come to have a distinctively weaker interaction with the elements 
in the `solid'-like phase.
Such differentiation is indispensable to the
storage of a large amount of energy.  
Partial contact with the heat bath is essential 
to maintain the differentiation of states needed to store the absorbed energy.
Our coupled pendulum model provides the simplest example for this behavior.

The dynamic mechanism for energy storage presented here
is simple enough to be realized in real protein motors.
It is only necessary that the protein dynamics  
exists near a phase transition region with `solid'-like (vibrational) modes
and `gas'-like (rotational) modes
and that the folded structure prevents some part of protein 
from experiencing the random effect of the heat bath.
These conditions are expected to be satisfied for many kinds of proteins 
with a sufficiently large size, in addition to the protein motors.

The authors are grateful to
T. Yanagida, Y. Ishii and members of the Single Molecule Processes Project, JST
for many meaningful suggestions.
They would also like to thank
T. Yomo, T. S. Komatsu, S. Sasa, and T. Shibata
for stimulating discussions.  
This research was partially supported by Grants-in-Aids for Scientific 
Research from the Ministry of Education, Science and Culture of Japan.

\begin{figure}[h]
\caption{Large deviation of the total energy $E=K+V$ at thermal equilibrium,
accompanied by the occasional appearance of the rotational mode
(See the region $50000<$time$<60000$).
This behavior is characteristic of the medium temperature.
For $i_c\ll N$, the rotational mode is mainly taken by pendula
without direct contact to the heat bath.
$T=0.02$, $N=10$, $i_c=1$ and $\gamma=10^{-2}$.
}
\label{fig:eng_abs}
\end{figure}

\begin{figure}[h]
\caption{Process of energy storage at $T=0.02$ in the case of 
partial contact with the heat bath ($i_c=1$ with $N=10$).
The values of the parameters are $\gamma=10^{-2}$, $\beta=10^{-3}$ and $V_0=2$.
(a) Time series of the momenta $p_i$.
The $N$th pendulum, kicked at $t=0$, reaches thermal equilibrium at $t_{R0}$,
as indicted by A.
The excess energy is transferred to another pendulum ($i\neq 1,N$),
as shown by B.
(b) The time series of the total energy $E$.
The inset is a close-up of the graph for time$>10000$.
}
\label{fig:eng_str}
\end{figure}

\begin{figure}[h]
\caption{Lifetime for energy storage when
$N=10$, $\Gamma=10^{-3}$, $\beta=10^{-2}$ and $V_0=4$.
(a) Distributions of the lifetime $\log_{10}(t_R-t_{R0})$
for energy storage at $T=0.02$ (medium temperature).
Each distribution is obtained from $200$ samples.
The unshaded distribution is that for the case of 
partial contact ($i_c=1$ with $\gamma=10^{-2}$), while the
shaded distribution is that for full contact ($i_c=10$ with $\gamma=10^{-3}$).
(b) Dependence of the mean lifetime $\langle t_R-t_{R0}\rangle$ 
for energy storage on the temperature $T$.
The values are averaged through $200$ samples.
The solid line denotes the case of partial contact 
($i_c=1$ with $\gamma=10^{-2}$),
while the dotted line denotes that for 
full contact ($i_c=10$ with $\gamma=10^{-3}$).
}
\label{fig:lifetime}
\end{figure}

\end{document}